# Proactive Decision Support using Automated Planning

Satya Gautam Vadlamudi, Tathagata Chakraborti, Yu Zhang, Subbarao Kambhampati
{gautam,tchakra2,Yu.Zhang.442,rao}@asu.edu, Arizona State University, Tempe, AZ

Proactive decision support (PDS) helps in improving the decision making experience of human decision makers in human-in-the-loop planning environments. Here both the quality of the decisions and the ease of making them are enhanced. In this regard, we propose a PDS framework, named RADAR, based on the research in Automated Planning in AI, that aids the human decision maker with her plan to achieve her goals by providing alerts on: whether such a plan can succeed at all, whether there exist any resource constraints that may foil her plan, etc. This is achieved by generating and analyzing the landmarks that must be accomplished by any successful plan on the way to achieving the goals. Note that, this approach also supports naturalistic decision making which is being acknowledged as a necessary element in proactive decision support, since it only aids the human decision maker through suggestions and alerts rather than enforcing fixed plans or decisions. We demonstrate the utility of the proposed framework through search-and-rescue examples in a fire-fighting domain.

Human-in-the-loop planning (HILP) is a necessary requirement today in many complex decision making or planning environments. In this paper we consider the case of HILP where the human(s) responsible for making the decisions in complex scenarios are supported by automated planning systems. Thus the planners in this scenario are the humans themselves, and we investigate the role of an automated planner in their deliberative process. This is, in effect, a role reversal of the traditional notion of the human-planner interaction in mixed initiative planning; and we refer to the proposed system as a reverse mixed initiative planner. These systems are capable of providing plans or course-of-actions (COAs) when a model of the world where the plans are to be executed is given to them, along with the knowledge of the initial state and the list of goals to be achieved/tasks to be accomplished. Examples where such technologies can be helpful include disaster response strategies from the navy, or responses to fire or emergency from local law enforcement.

Providing a complete model of the world where the plans are to be executed is, however, known to be very difficult (Kambhampati, 2007). This implies that the system generated plan cannot be completely relied upon. Not only executing such plans may no longer accomplish the goals/tasks provided, but also their execution may result in undesired consequences. This calls for active participation from the human in the loop rather than simply adopt a system generated plan. Furthermore, in many cases, the human in the loop may be held responsible for the plan under execution and its results. Therefore, it is also necessary in such cases that the human keeps control of the plan/COA being given for execution. This motivates us to build a proactive decision support system, which is context-sensitive and focuses on aiding and alerting the human in the loop with his/her decisions rather than generate a static COA that may not work in the dynamic worlds that the plan has to execute in.

In this paper, we propose a proactive decision support (PDS) system, named RADAR, using automated planning technology, which is augmentable, context sensitive, controllable and adaptive to the human's decisions. It supports the human in the loop through suggestions and alerts, which can be considered by the human as he/she sees fit.

In the following we explain the meaning of the above terms:

- **Augmentable**: The model of the world such as the rules that specify what are the preconditions for a particular action/decision and what would be the effects of that action/decision could be added/modified. The state of the world such as the values of various variables and availabilities of various resources could be updated. The list of goals/tasks can also be updated.
- **Context-sensitive**: Whenever the model or the state of the world is augmented either by the human or by some other source internal/external to the PDS system, the system takes the new context into account and responds with updated suggestions and alerts.
- **Controllable**: The decision making process of the PDS system is completely controllable by the human in the loop, who retains with the decision making power wherein he/she can either choose to follow the system generated suggestions or make a different decision as they see fit.
- **Adaptive**: Since the decision making power lies with the human in the loop, the PDS system has to adapt itself to the decisions being made by the human and provide new suggestions and alerts that are relevant based on the decisions of the humans. Note that, adaptive nature may also be viewed as part of context-sensitivity in the sense that the context changes whenever decisions are made. In this paper, we keep the distinction to differentiate the changes in the world and tasks, and the changes related to the actions prescribed/decisions of the human in the loop.

As mentioned before, our proactive decision support system uses automated planning technology widely studied in the field of Artificial Intelligence. In particular, we adopted the Planning Domain Description Language (PDDL) to describe the model of the world, details of the current state (context) and the goals, to the system. Then, we used the existing landmark generation method (Hoffmann et al., 2004) to generate landmarks, which are then analyzed to come up with relevant suggestions and alerts. Landmarks are those set of states (of the world) that has to be visited by any successful plan that achieves goals from the current state.

We implemented our proposed system using the Fast Downward planner (Helmert, 2006) and tested it on a fire-fighting domain, where search-and-rescue missions are to be carried out (goals). We also conducted some preliminary human factor studies to evaluate the utility of the above proposal, which gave positive feedback.

## RELATED WORK

The proposed proactive decision support system supports naturalistic decision making (Zsambok and Klein, 2014; Klein, 2008), which is acknowledged as a necessary element in PDS systems (Morrison et al., 2013). Systems which do not support naturalistic decision making have been found to have detrimental impact on work flow causing frustration to decision makers (Feigh et al., 2007).

In (Parasuraman, 2000), a study of human performance consequences of different levels and types of automation is provided, where aspects such as, mental workload and situation awareness are considered as evaluative criteria. A model for types and levels of automation that provides an objective basis for deciding which system functions should be automated and to what extent is given in (Parasuraman et al., 2000). (Parasuraman and Manzey, 2010) shows that human use of automation may result in automation bias leading to omission and commission errors, which underlines the importance of reliability of the automation (Parasuraman and Riley, 1997). Various elements of human-automation interaction such as, adaptive nature and context sensitivity are presented in (Sheridan and Parasuraman, 2005). (Warm et al., 2008) show that vigilance requires hard mental work and is stressful via converging evidence from behavioral, neural and subjective measures. Our system could be considered as a part of such vigilance support thereby reducing the stress for human in the loop.

High-level information fusion that characterizes complex situations and that support planning of effective responses is considered the greatest need in crisis-response situations (Laskey et al., 2016). Automated planning based proactive support systems were shown to be preferred by humans in studies involving human-robot teaming (Zhang et al., 2015) and the cognitive load of the subjects involved was observed to have been reduced (Narayanan et al., 2015).

## PROPOSED PROACTIVE DECISION SUPPORT SYSTEM: RADAR

Now, we present the proposed proactive decision support (PDS) system – RADAR, based on automated planning technology. First, we present the different elements of the PDS system and then briefly present the details of the methodology behind the generation of suggestions and alerts. The various elements of the planning PDS system are:

- *Tasks/goals:* The tasks or goals to be accomplished clearly form an important and necessary element.
- *State/context, resources:* The current state or context is needed for the PDS system to produce relevant alerts. The availability of resources also forms part of the context which we separately indicated to display as an important constituent of the Context.
- *Model, actions/decisions:* The model consists of set of rules which are applicable in the world where the plan is being executed. Actions, for example, are part of a model, which give information about when a particular action is applicable (what are the pre-conditions to be satisfied in order for it to be applicable), and what would be the effects of taking that action (how it would impact various elements of the world). Actions are also closely tied to decisions that need to be made since each decision typically corresponds to certain action being taken.
- *Current plan/course of action (COA):* The information about current plan of the human in the loop, if any, can help the PDS system produce better suggestions and alerts by reducing the uncertainty. However, this could just consist of actions already taken, in which case the proposed PDS system can come up with relevant alerts pertaining to the future. More details on this are given next.

Now, we briefly present details on how the proposed planning based system with above elements produces relevant suggestions and alerts. In order to explain this, first we need to define what are called Landmarks, which are central to the suggestions and alerts system.

*Definition:* **Landmarks.** (Hoffmann et al., 2004) A state/partial state is a landmark (for the current state, tasks/goals, and model) if all plans/course-of-actions that can accomplish the tasks from the current state must go through that state/partial state during their execution.

Note that, all goal states are trivial landmarks since they have to be accomplished by all successful plans. Consider there exists only one state, **A**, which can take one to the goal state(s), meaning, all plans have to visit **A** in order to accomplish the goals, making it a landmark (derived; non-trivial). Further, if there exist two states **A** and **B** through which the goal state(s) must be reached, then either **A** or **B** must be visited before accomplishing the tasks/goals. In such a scenario, **A or B** can be called as a landmark, in particular, a **disjunctive landmark.** Continuing this process, one can derive recursively the set of all landmarks starting from the goal state(s) leading back to the current state.

**Generating Suggestions and Alerts (PDS)**

Now, we present the details on how generating and analyzing the landmarks can help in producing suggestions and alerts.

Note that, since the landmarks are the states that must be visited in order to accomplish the goals, if there is no possible way of reaching a certain landmark generated above, then the system can generate an *alert* conveying that the goal cannot be accomplished. This could be because there is no action available, which would help in visiting the landmark under consideration, or the preconditions of actions that can help reaching the landmark are not satisfied.

In some cases, the preconditions, which are not currently satisfied for an action to be applicable, may be because of resource constraints. In such cases, the system instead generates a *suggestion* mentioning that those resources are needed in order to accomplish the task. For example, in order for an action such as *put-off-fire* to be applicable, a precondition on the availability of the resource: *fire-engine* need to be satisfied. Failing which an alert may be generated. More details are described through a case study next.

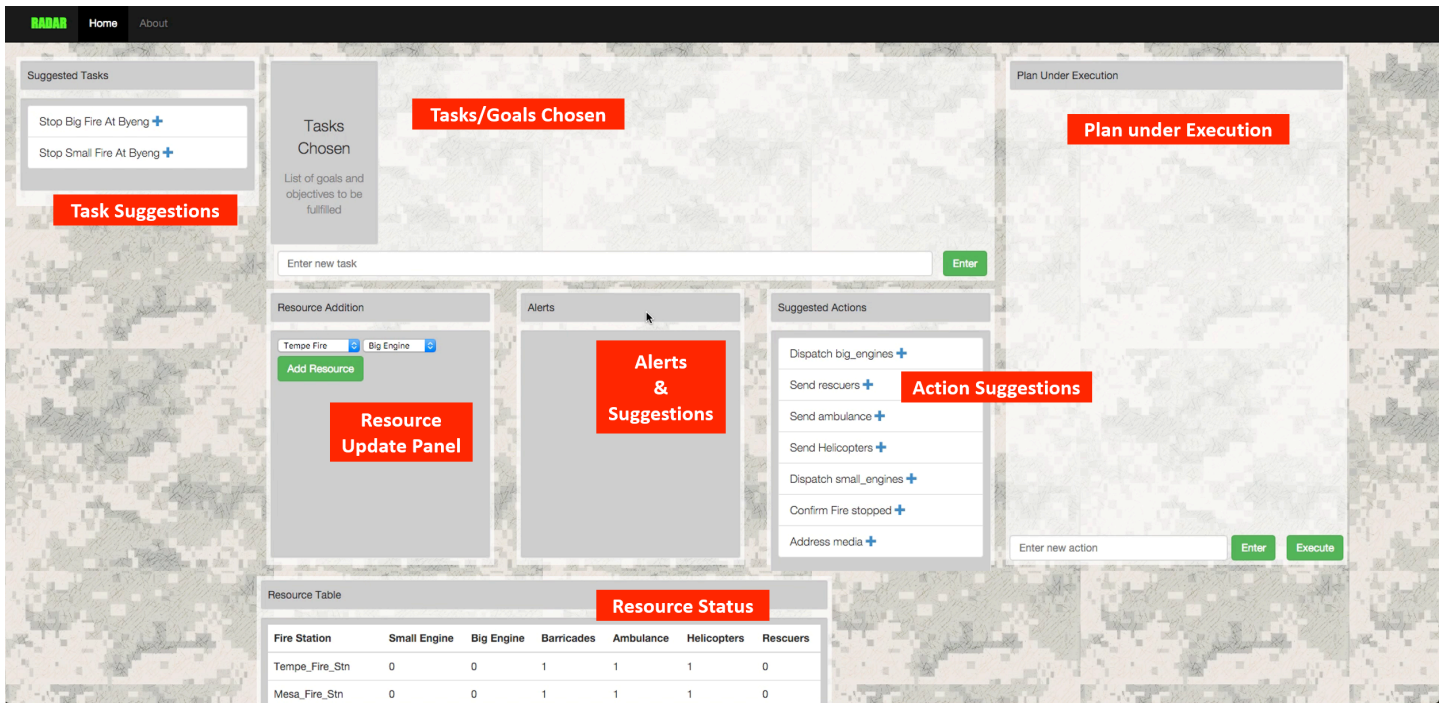

Figure 1 – RADAR interface showing data support and decision support for the human commanders making plans.

Further, provision to update the model, state, and resources is provided to make the system augmentable. In order to support context-sensitivity and adaptive nature, we re-execute the alerts generation method whenever there is a change in the context/tasks or new action is executed or plan is changed, so that the suggestions/alerts become relevant to the situation.

**Case Study: Fire-fighting Domain**

We use a fire-fighting scenario to illustrate the ideas expressed so far, as shown in Figure 1. The scenario plays out in a particular location (we use Tempe in the example) and involves the local fire chief, police, medical and transport authorities, who try to build a plan in response to the fire in the given platform (which is augmented with decision support capabilities from an automated planner). The left pane gives event updates that the commanders can incorporate into the Tasks panel at the top, which shows what high level goals or tasks needs to be addressed. The panel on the right (currently empty) will display the plan being constructed. Each of the human commanders have access to the resources that they can use to control the fire outbreak (as can be seen from the table at the bottom of Figure 1). For example, the police can deploy police cars and policemen, and the fire chief can deploy fire engines, ladders, rescuers, etc.

The plans that can be produced by the commanders, of course, depend on the availability of these resources, and certain actions can only be executed when the required number of resources are available or the preconditions are satisfied. For example, in order to dispatch police cars from a particular police station, the police chief needs to make sure that the respective police station has enough police cars and it has been notified of the demand previously.

Given this knowledge, the automated planner integrated into the system keeps an eye on the planning process of the human commanders. The three panels in the middle of Figure 1 provide these functionalities. The one on the left provides a way to add or request for resources in case of insufficient resources, while the one on the right provides suggested actions that the commanders may use to complete their plans. The panel in the middle is the most important part of the automated component where it produces alerts or suggestions to problems in the current plan or with respect to problems that may appear in future given the current state and availability of resources.

In the following we will go through two use cases to illustrate how the system responds to situations as per the guidelines we discussed in the introduction –

**Scenario I: (see Figure 2)**

1) The scenario starts with a small fire in a building. Once selected from suggested tasks, this populates the task chosen to be addressed.
2) The planning model does landmark analysis on the current state and immediately populates the alerts panel with an alert saying either big fire engines or small fire engines are needed to put out the fire (a disjunctive landmark).
3) The commander tries to dispatch big engines now, but is stopped by the system which detects that there are not enough resources (big engines).
4) The commander addresses the shortage by requesting additional engines from the left.
5) Now the commander can proceed with and finish the plan until the fire has been extinguished.

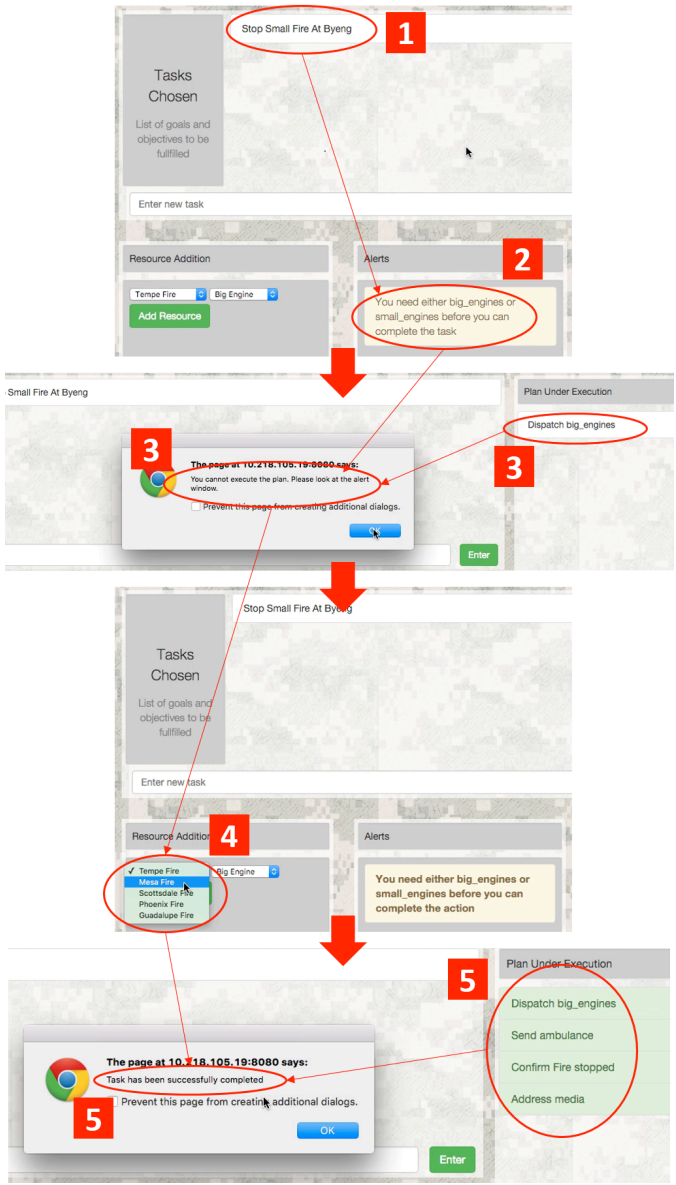

Figure 2 – Use case illustrating automated decision support using disjunctive landmarks.

**Scenario II: (see Figure 3)**

1) The scenario now starts with a big fire. This calls for big engines as alerted by the system. Note that it also alerts for insufficient rescuers, which did not happen in the previous case, as the model used conveys that rescuers are needed only in case of a big fire.
2) The commander once again requests for additional resources (big engines as well as rescuers) and was able to generate a feasible plan as shown.

In this way the system is able to assist the human commander in his planning process. The system is *augmentable* in the way it supports the commander's goal preferences and world state information. It is *context-sensitive* in how it provides relevant alerts based on the stage of the plan, and *adaptive* with respect to the choice taken by the human in the loop to address these alerts. Finally, the entire process is *controllable* because the human commander has authority over the choices at all times.

Note that the above examples are illustrative and only intended for understanding the underlying concepts of the proposed system. The same system can handle scenarios where hundreds of actions and variables are involved, where it becomes nearly impossible for a human to account for all possible drawbacks. Furthermore, any other domain (say, disaster response) can be readily handled by the implemented system, by just changing the PDDL domain file used by it without any renewed effort.

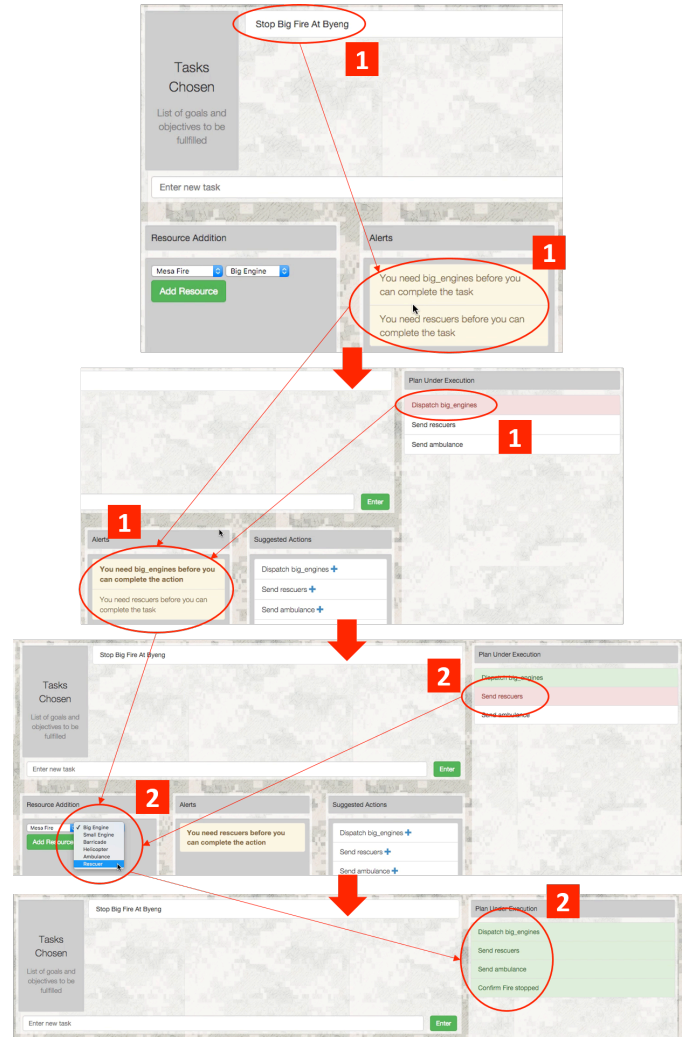

Figure 3 – Use case illustrating how automated decision support adapts in response to a necessary landmark.

**EVALUATIONS WITH HUMANS ON UTILITY**

In order to assess the utility of the proposed proactive decision support (PDS) system, RADAR, we have conducted a preliminary survey of its usefulness on 7 subjects on the questions given below. The system would be enhanced and refined continuously as we incorporate more and more features in our PDS system. We also plan to take the learning from the user feedback back into the design of the PDS system during the process. We discuss the responses obtained for each of the questions, next to it.

*1) Do you think the suggestions & alerts are relevant to the task/goal? Yes/No*

7 out of 7 subjects have answered it Yes. This suggests that the domain shown (fire-fighting) has been modeled well so that relevant landmarks could be generated.

*2) Do you think the suggestions & alerts are context-sensitive (current state & current plan)? Yes/No*

7 out of 7 subjects have answered it Yes. This suggests that the context sensitivity of the landmarks make a good fit for use in PDS systems.

*3) Do you think that the suggestions & alerts are dynamic (change with changing context/plan)? Yes/No*

7 out of 7 subjects have answered it Yes. This suggests that the regeneration of suggestions & alerts whenever context changes is notable to the users.

*4) Do you think the suggestions & alerts increased your situational awareness (e.g. resources available, status of execution)? Yes/No*

7 out of 7 subjects have answered it Yes. This suggests that the suggestions and alerts continuously improve the situational awareness of the human in the loop, particularly in terms of the critical points relevant to the plan/COA.

*5) Do you think that the suggestions & alerts interfere with the ability to interact with the system? Yes/No*

1 out of 7 subjects have answered it Yes. This suggests that the alerts are displayed without interfering with the user interaction experience in most cases, and may be improved.

*6) Do you think the suggestions & alerts helped in debugging the plan or would you rather execute and replan in case of failure? Former/Latter*

6 out of 7 subjects have answered that the suggestions and alerts helped. This suggests that the users prefer to be alerted in advance rather than re-plan in most of the cases.

*7) Do you think the suggestions & alerts should be an error and stop the plan from being dispatched, or should it be a warning and let you proceed with execution? Former/Latter*

4 out of 7 subjects have answered that the alerts can be shown as errors and stop the execution whereas others preferred them as warnings that allow execution to move forward. In this case, there is a split amongst the users as to whether the user should be able to proceed with warnings or be stopped completely. Here there is an opportunity to amend the system so as to learn the cases where one could proceed despite the error/warning, and incorporate it as a soft constraint in the future leading to only a warning.

*8) Do you feel that you can be in control of the decision making/planning process when using the proposed PDS system? Yes/No*

6 out of 7 subjects have answered it Yes. This suggests that most of the users feel in control of the decision making process rather than being forced by the system.

*9) On a scale of 1-10 how much would you rate your satisfaction with the proactive decision support capabilities of the system? 1-10*

This received an average score of 7.14. This suggests that the users had a good first experience with the proposed PDS system and would like to see new and improved features.

*10) On a scale of 1-10 how much would you recommend using the proposed proactive decision support platform? 1-10*

This received an average score of 7.28. This suggests that the users are open to recommending the PDS system to others.

Overall, the PDS system was well received and perceived to be promising. Users have appreciated the current features and suggested minor modifications. As part of the future work, we consider addressing several important aspects such as: Where do we get the models? Can we automatically learn them from observing the users and the contexts? How do we deal with incomplete models? Does the human in the loop deviate from cost-optimal plans? How to understand the preferences of the human in the loop and how to address them? And so on.

## CONCLUSION

We presented a Proactive decision support (PDS) framework called RADAR, based on the research in Automated Planning in AI, that aids the human decision maker with her plan to achieve her goals by providing alerts and suggestions on potential drawbacks in the plan and resource constraints. This was achieved by generating and analyzing the landmarks that must be accomplished by any successful plan before achieving the goals. The proposed approach is aligned with the concept of naturalistic decision making. We demonstrated the utility of the proposed framework through search-and-rescue examples in a fire-fighting domain and human factors studies.

## ACKNOWLEDGMENTS

This research is supported in part by the ONR grant N00014-15-1-2027. We thank Vivek Dondeti for his help with the implementation of parts of the RADAR system.